\documentclass[aps,prl,twocolumn,showpacs,superscriptaddress,floatfix,citeautoscript]{revtex4}
\usepackage{dcolumn}
\usepackage{bm}
\usepackage{amsmath,amssymb}
\usepackage{times}
\usepackage{float}
\usepackage{graphicx, epstopdf}

\begin{document}

\title{High Contrast X-ray Speckle from Atomic-Scale Order in Liquids and Glasses}

\author{S. O. Hruszkewycz}
\affiliation{Materials Science Division, Argonne National Laboratory, Argonne, Illinois 60439 USA}
\author{M. Sutton}
\affiliation{Department of Physics, McGill University, Montreal, H3A2T8 Canada}
\author{P. H. Fuoss}
\affiliation{Materials Science Division, Argonne National Laboratory, Argonne, Illinois 60439 USA}
\author{B. Adams}
\affiliation{X-ray Science Division, Argonne National Laboratory, Argonne, Illinois 60439 USA}
\author{S. Rosenkranz}
\affiliation{Materials Science Division, Argonne National Laboratory, Argonne, Illinois 60439 USA}
\author{K. F. Ludwig~Jr.}
\affiliation{Department of Physics, Boston University, Boston, Massachusetts, 02215 USA}
\author{W. Roseker}
\affiliation{HASYLAB, Deutsches Elektronen-Synchrotron, Hamburg, Germany}
\author{D. Fritz}
\affiliation{Linac Coherent Light Source, SLAC National Accelerator Laboratory, Menlo Park, California 94025 USA}
\author{M. Cammarata}
\affiliation{Linac Coherent Light Source, SLAC National Accelerator Laboratory, Menlo Park, California 94025 USA}
\author{D. Zhu}
\affiliation{Linac Coherent Light Source, SLAC National Accelerator Laboratory, Menlo Park, California 94025 USA}
\author{S. Lee}
\altaffiliation{Present address: Physical Metrology Division, Korea Research Institute of Standards and Science, Daejeon 305-340 S. Korea.}
\affiliation{HASYLAB, Deutsches Elektronen-Synchrotron, Hamburg, Germany}
\affiliation{Linac Coherent Light Source, SLAC National Accelerator Laboratory, Menlo Park, California 94025 USA}
\author{H. Lemke}
\affiliation{Linac Coherent Light Source, SLAC National Accelerator Laboratory, Menlo Park, California 94025 USA}
\author{C. Gutt}
\affiliation{HASYLAB, Deutsches Elektronen-Synchrotron, Hamburg, Germany}
\author{A. Robert}
\affiliation{Linac Coherent Light Source, SLAC National Accelerator Laboratory, Menlo Park, California 94025 USA}
\author{G. Gr\"{u}bel}
\affiliation{HASYLAB, Deutsches Elektronen-Synchrotron, Hamburg, Germany}
\author{G. B. Stephenson}
\affiliation{Materials Science Division, Argonne National Laboratory, Argonne, Illinois 60439 USA}
\affiliation{Advanced Photon Source, Argonne National Laboratory, Argonne, Illinois 60439 USA}

\date{\today}

\begin{abstract}  
The availability of ultrafast pulses of coherent hard x-rays from the Linac Coherent Light Source
opens new opportunities for studies of atomic-scale dynamics in amorphous materials.
Here we show that single ultrafast coherent x-ray pulses can be used to observe the speckle contrast
in the high-angle diffraction from liquid Ga and glassy Ni$_2$Pd$_2$P and B$_2$O$_3$.
We determine the thresholds above which the x-ray pulses disturb the atomic arrangements.
Furthermore, high contrast speckle is observed in scattering patterns 
from the glasses integrated over many pulses,  
demonstrating that the source and optics are sufficiently stable for 
x-ray photon correlation spectroscopy studies of dynamics over a wide range of time scales.
\end{abstract}


\maketitle

Ever since the initial observation of speckle 
in the diffraction of coherent short-wavelength x-rays \cite{SUTTON:1991p2169}, 
it has been recognized that coherent x-ray techniques have the potential to provide powerful new probes 
of {\it atomic scale} structure and dynamics in non-crystalline systems,
analogous to the techniques developed to study disorder at much larger length scales 
using coherent visible light \cite{BernePecora1976}.
In particular, x-ray photon correlation spectroscopy (XPCS) has been developed 
using high-brightness synchrotron x-ray sources 
to observe equilibrium and non-equilibrium dynamics in diverse systems \cite{LivetAC2007, StephensonNM2009}.
A limiting factor in the development of XPCS for atomic scale studies has been the signal level that can be obtained.
Almost all XPCS studies to date probe relatively large length scale ($> 10$~nm) structures,
for example via small-angle scattering from colloids and 
polymers \cite{DierkerPRL1995, RiesePRL2000, PontoniPRL2003, MochriePRL1997, LummaPRL2001}
or diffuse scattering near Bragg peaks of crystals \cite{Brauer1995, Shpyrko:2007p3209, Pierce:2009p3941},
because of their larger scattering cross sections compared with atomic-scale fluctuations,
and the limited coherent x-ray power available. 
The first XPCS observation of truly atomic-scale dynamics was obtained 
only recently \cite{Leitner:2009p4799} in a study of diffusion in Cu-Au on relatively slow timescales ($> 10$~s).

One of the scientific drivers for the new generation of intense, coherent hard x-ray free electron laser sources, 
such as the Linac Coherent Light Source (LCLS) at SLAC National Accelerator Laboratory, 
has been their potential for XPCS studies to open a new frontier at the natural time scales 
of even the fastest condensed matter systems, e.g. atomic diffusion in liquids \cite{First_Experiments, GruebelNIM2007}.
The combination of femtosecond pulses and atomic resolution speckle 
provides unprecedented opportunities to test theories of liquid structure and dynamics, 
such as the intriguing predictions that have recently emerged regarding the complex dynamics 
of both diffusive \cite{Langer2008} and vibrational \cite{Shintani2008} modes in liquids and glasses. 
Ultrafast XPCS provides a time-domain probe complementary to inelastic scattering, 
capable of studies of sub-micron volumes and non-equilibrium dynamics.
To reach femtosecond time scales, which are much faster than the time resolution of imaging detectors,
a pulse split-and-delay technique has been proposed \cite{First_Experiments, GruebelNIM2007, GuttOE2009, RosekerOL2009, RosekerJSR2011} 
in which the correlation time is determined from the change in contrast of the sum of two speckle patterns, 
as a function of the delay time between the patterns.
Although LCLS provides a huge leap in available coherent flux and accessible time scales, 
the high number of photons delivered in a single pulse poses challenges in avoiding x-ray effects 
on the sample during a  dynamics measurement \cite{First_Experiments, GruebelNIM2007}.
A primary issue is whether accurate contrast values can be extracted from the relatively weak speckle patterns 
expected in optimized XPCS studies. 
Likewise, the variations in pulse energy and position 
pose potential challenges for speckle measurements.
Here, we report the first observation of high contrast x-ray speckle 
from the atomic scale structure in both liquid and glass samples. 
These were obtained using femtosecond hard x-ray pulses from the LCLS. 
We present an effective analysis procedure to extract contrast, 
show the dependence of contrast on experimental conditions, 
report the observed perturbation thresholds of various samples, 
and discuss the design of an optimized atomic resolution XPCS experiment.

We recorded speckle patterns from three samples:
Ga heated into the liquid phase at 35 C,
and two glasses at room temperature, Ni$_2$Pd$_2$P and B$_2$O$_3$.
The average structure factors of these systems have all been previously characterized
using standard scattering methods \cite{Ga, NiPdP, B2O3}.
Experiments were carried out with the XPP instrument at LCLS,
using a Si (111) monochromator
with $1.4 \times 10^{-4}$ bandwidth set to 7.99 keV photon energy,
and a $70$ fs electron pulse width \cite{Supplemental}.
The average unattenuated x-ray pulse energy was typically  
$\sim 1 \times 10^{10}$ and $\sim 4 \times 10^{10}$ photons per pulse
for single-pulse or multi-pulse images, respectively \cite{Supplemental},
although there was a very broad distribution of pulse energies
with a maximum typically 5 times the mean.
The scattering geometries were chosen to provide
coherent illumination conditions that would fulfill the requirements to produce
high-contrast speckle patterns \cite{PuseyReview, SuttonReview}
and to resolve this pattern at the detector. 
Measurements were made with a beam focused to $1.7$ $\mu$m FWHM
at the sample using Be compound refractive lenses.
Diffraction patterns were recorded using a direct-x-ray-detection CCD with
20 $\mu$m square pixels at a distance from the sample of $L = 18$ to 142 cm,
positioned at an average scattering angle corresponding to the peak in the strongest amorphous scattering ring.
For Ga and Ni$_2$Pd$_2$P, we employed bulk samples in reflection geometry at
$Q \approx$ 2.60 and 2.95 \AA$^{-1}$, respectively;
for B$_2$O$_3$, we employed a fiber of 18 $\mu$m thickness in transmission geometry
at $Q \approx$ 1.62  \AA$^{-1}$.
Unattenuated single-pulse speckle patterns were recorded from all samples.
To investigate the stability of the experimental setup and perturbation of the sample by the x-ray beam,
we also recorded speckle patterns from the glass samples 
that were sums of multiple attenuated pulses.

Figure 1(a) shows a typical region of amorphous scattering viewed by the detector
at $L = 37$ cm for liquid Ga.
This image of the average scattering was obtained by summing many single-pulse patterns
and binning the pixels to smooth out the photon statistics.
Figure 1(c) gives a small region of one of the single-pulse patterns, showing the individual photon hits.
Since the width of the x-ray pulses is expected to be shorter than 
the time scale of equilibrium diffusive and vibrational motions of the atoms in the liquid,
such single-pulse diffraction patterns should exhibit high speckle contrast
if the x-ray optics are sufficiently free of aberrations, the detector resolution and noise characteristics are adequate,
and the x-ray pulse does not disturb the atomic positions in the sample on the time scale of the measurement.
A critical issue is to understand the accuracy with which the contrast can be determined
from such weak speckle patterns.

Recent studies have obtained the contrast in speckle patterns
from the normalized variance of the signal distribution on the detector \cite{GuttPRL2011,LudwigJSR2011}.
When the photon density is low,
the variance contains a large contribution from the counting statistics of individual photons
in addition to the contrast \cite{GuttOE2009,LivetAC2007, Supplemental}.
Our approach to extracting the contrast 
is to compare the observed probability for $k$ photons arriving in a single detector pixel 
to the negative binomial distribution expected for a speckle pattern of a given contrast \cite{NB}
\begin{equation}
P(k) = \frac{\Gamma (k+M)}{\Gamma (M) \Gamma (k+1)} \left(1+\frac{M}{\bar{k}}\right)^{-k} \left(1+\frac{\bar{k}}{M}\right)^{-M},
\end{equation}
where $\bar{k}$ is the mean photon density (counts per pixel), 
$M$ is the number of modes in the speckle pattern, and $\Gamma$ is the gamma function.
The speckle contrast factor $\beta \equiv M^{-1}$ 
has a maximum value of unity
for a single-mode distribution,
and it drops to zero as $M$ increases.
For $M \rightarrow \infty$, 
Eq.~(1) approaches the familiar Poisson distribution for uncorrelated photon positions, 
$P(k) = \bar{k}^k \exp(-\bar{k}) / k!$.
The contrast factor will be less than unity if the
conditions are not met  for the required transverse or longitudinal coherence 
of the incident beam \cite{PuseyReview, SuttonReview}, 
if the detector pixels are too large to fully resolve the speckle, 
or if the sample is not static on the time scale over which the pattern is recorded.
It is this latter effect that allows the sample dynamics to be extracted 
from measurements of speckle contrast.
In particular, if we record the sum of two speckle patterns of contrast factor $\beta_0$ 
from two equal-energy x-ray pulses separated by a time $\tau$,
the contrast factor of the sum will drop from $\beta_0$ to $\beta_0/2$
as $\tau$ is varied from much less than to much more than the correlation time 
of the sample dynamics \cite{GruebelNIM2007,NB}.

\begin{figure}
\includegraphics[width=3.0in]{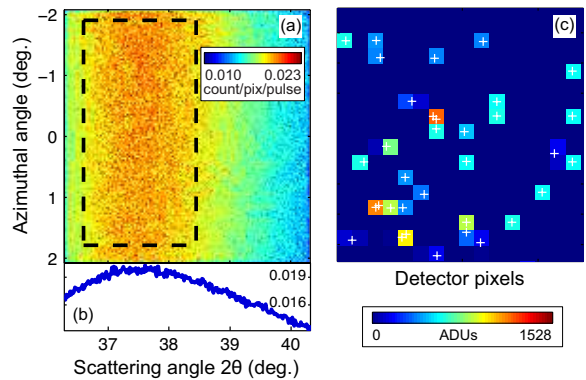}
\caption{(Color online) (a) Average scattering of liquid Ga integrated over 309 pulses from LCLS.
Detector at $L = 37$ cm from sample subtends scattering angles
including the main amorphous scattering ring at $Q = 2.60$~\AA$^{-1}$.
Dashed box shows region of pixels used in contrast analysis.
(b) Average photon density vs. scattering angle.
(c) Expansion of small region of a single-pulse scattering pattern showing 
signal in analog-to-digital units (ADUs) 
from individual photons, and assignment of photon positions.}
\vspace{-0.1 in}
\end{figure}

To accurately extract the contrast in these weak speckle patterns,
we first employ a ``droplet algorithm'' \cite{Supplemental,Livet:2000p4334}
to locate the positions of each detected photon
by fitting each recorded pattern.
Typical results for extracted photon positions are shown in Fig. 1(c).
Experimental probabilities $P(k)$ are obtained by choosing a region of pixels
that has nearly uniform $\bar{k}$ and $Q$,
i.e. along the peak of the amorphous scattering ring as shown in Fig. 1(a),
binning the photon positions into the detector pixels,
and determining the fraction of pixels that have $k$ photons.
Figure 2 shows the experimental probabilities for $k = 1$ to 4
for 309 speckle patterns having various mean count densities $\bar{k}$.
We obtained a contrast factor $\beta$ for each speckle pattern
by analyzing the $P(k)$ as described in the supplemental material  \cite{Supplemental}.
The weighted average of these values
gives $\langle \beta \rangle = 0.276 \pm 0.004$
for liquid Ga and $L = 37$ cm.
While the speckle patterns with highest $\bar{k}$
provide the highest accuracy contrast values,
the full distribution of weaker and stronger speckle patterns
obtained (due to the wide incident pulse energy variation
characteristic of monochromatic hard x-ray experiments at LCLS) verifies
the predicted $\bar{k}$ dependences
given by the negative binomial distribution.
These measurements on liquid Ga represent the first observation of speckle 
from the atomic-scale order in a liquid.

\begin{figure}
\includegraphics[width=3.0in]{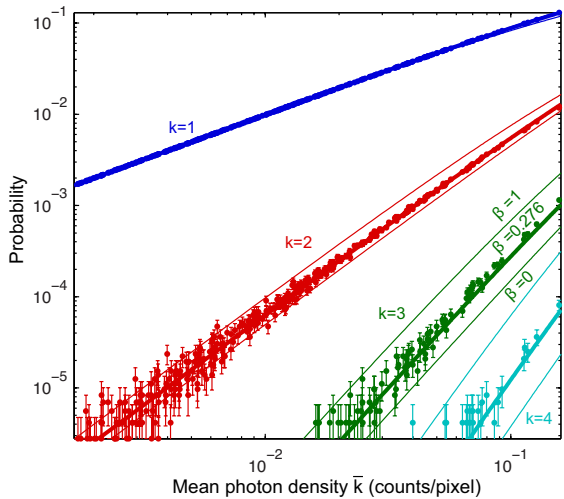}
\caption{(Color online) Observed probability of $k = 1$ to 4 photons within a pixel 
as a function of mean photon density $\bar{k}$, 
for 309 single-pulse speckle patterns from liquid Ga for $L = 37$ cm.
Thick curve is negative binomial distribution
with $\beta = 0.276$;
thin curves show limiting distributions
for $\beta = 1$ and $\beta = 0$.}
\vspace{-0.1 in}
\end{figure}

\begin{table}
\caption{Experimental parameters, extracted average contrast factors $\langle \beta \rangle$,
and calculated maximum contrast factor $\beta_{calc}$ for speckle patterns from
Ga liquid, Ni$_2$Pd$_2$P glass, and B$_2$O$_3$ glass.
The parameter $L$ is the sample-to-detector distance,
``pulses'' is the number of consecutive LCLS pulses integrated in each speckle pattern,
$n_{patt}$ gives the the number of speckle patterns included in the contrast analysis,
$\langle I_0 \rangle$ gives the average number of photons per pulse incident on the sample (after any attenuation),
and $\langle \bar{k} \rangle$ is the average of the mean photon densities in this set of patterns.
The number of pixels used was $n_{pix} = 3.6, 7.2, \rm{or}~ 13.2 \times 10^5$
for $L = 18, 35 ~\rm{to}~ 37, \rm{or}~ 142$ cm, respectively.
}
\label{tabsumm}
\begin{ruledtabular}
\begin{tabular}{r|r|r|l|l|l|l}
$L$ & pulses & $n_{patt}$ & $\langle I_0 \rangle$ & $\langle \bar{k} \rangle$ & avg. contrast & $ 
\beta_{calc}$ \\
(cm)&   & & (ct/pulse)  & (ct/pix) & factor $\langle \beta \rangle$&\\
\hline
\bf{Ga\hspace*{0.5cm}} \\
37  &     1 & 309 & $1.1 \times 10^{10}$& 0.019 & $0.276 \pm 0.004$ & 0.307 \\ 
\hline
\bf{Ni$_2$Pd$_2$P} \\
18  &     1 & 29 & $1.5 \times 10^{10}$  & 0.072 & $0.245 \pm 0.006$  & 0.349 \\ 
18  & 500 & 60 & $1.7 \times 10^{7}$  & 0.065 & $0.276 \pm 0.005$ & 0.349 \\ 
18  & 500 & 47 & $8.4 \times 10^{7}$    & 0.184 & $0.156 \pm 0.002$ & 0.349 \\ 
35  &     1 & 81 & $1.2 \times 10^{10}$   & 0.016 & $0.360 \pm 0.010$ & 0.645 \\ 
35  & 500 & 40 & $1.8 \times 10^{7}$  & 0.014 & $0.492 \pm 0.022$ & 0.645 \\ 
35  & 500  & 60 & $8.7 \times 10^{7}$  & 0.041 & $0.319 \pm 0.006$ & 0.645 \\ 
142 & 500 & 20 & $4.4 \times 10^{8}$  & 0.016 & $0.408 \pm 0.019$  & 0.713 \\  
\hline
\bf{B$_2$O$_3$\hspace*{0.2cm}} \\
37  &   10 & 9 & $8.6 \times 10^{9}$   & 0.007 & $0.291 \pm 0.057$ & 0.446 \\ 
37  & 100 & 20 & $1.9 \times 10^{10}$  & 0.056 & $0.195 \pm 0.005$ & 0.446 \\ 
37  & 500 & 61 & $3.6 \times 10^{8}$  & 0.017 & $0.356 \pm 0.014$ & 0.446 \\ 
\end{tabular}
\end{ruledtabular}
\end{table}

To investigate whether the stability of the source and experiment will affect the contrast
in multi-pulse XPCS experiments, 
and to understand the thresholds for irreversible perturbation of the sample structure by the
incident x-ray pulses, 
we studied two glasses where the liquid-like atomic arrangement should be static.
For the metallic glass Ni$_2$Pd$_2$P,
we collected not only
single-pulse speckle patterns without attenuation of the incident beam
(like we did with liquid Ga),
but also sums of 500 consecutive attenuated pulses.
The values of $\langle \beta \rangle$ extracted for both single-pulse
and attenuated 500-pulse speckle patterns
at three different values of sample-to-detector distance $L$
are given in Table I.
Scattering from the low-atomic-number glass B$_2$O$_3$
was too weak to obtain values of $\langle \beta \rangle$
from single pulses under the available experimental conditions;
Table I gives results for multiple-pulse patterns with various $\langle I_0 \rangle$.

The observed $\langle \beta \rangle$ values can be compared with those
calculated for a static sample
using an incident beam with full transverse coherence but non-zero photon energy bandwidth,
and a detector with non-zero pixel size.
These resolution effects reduce the calculated static contrast factor $\beta_{calc}$ 
from unity depending upon the size of the
illuminated volume, the scattering angle, 
and the sample-to-detector distance $L$ \cite{Supplemental}. 
Table I gives $\beta_{calc}$ values 
for each of the samples and experimental conditions.
The observed $\langle \beta \rangle$ value for liquid Ga is 90\% of the corresponding $\beta_{calc}$,
indicating that the ultrafast x-ray pulses have effectively frozen the motion of the atoms,
giving a coherent diffraction snapshot of their arrangement in the liquid.
It also indicates that the unattenuated, focused x-ray beam does not significantly
alter the atomic arrangements in liquid Ga during the time of the pulse,
even though the energy deposited in the illuminated sample volume
($\sim 30$ eV per atom per average pulse) is typically enough to vaporize it 
on a longer time scale \cite{Supplemental}.

\begin{figure}
\includegraphics[width=3.0in]{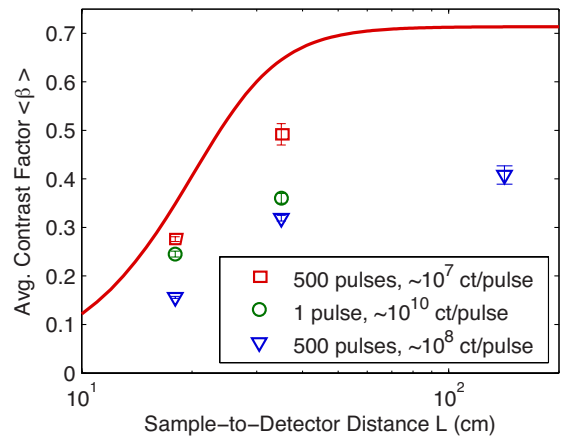}
\caption{(Color online) Contrast factor $\langle \beta \rangle$ vs. sample-to-detector distance $L$
for the Ni$_2$Pd$_2$P glass.
Curve is calculation for a static sample; 
symbols are observed values for single-pulse and 500-pulse images
with different $\langle I_0 \rangle$ values shown in legend.}
\vspace{-0.1 in}
\end{figure}

Figure 3 shows the measured contrast factor values for the Ni$_2$Pd$_2$P glass
compared with $\beta_{calc}$ as a function of $L$.
The overall dependence of the observed $\langle \beta \rangle$ on $L$ agrees with that of the calculation.
The highly attenuated 500-pulse patterns have contrast factors that are 75-80\% of $\beta_{calc}$,
indicating not only that there is minimal disturbance of the sample by the x-ray pulses,
but that the contrast is not greatly lowered by variations in the pulse energy and position of the incident beam
or other instabilities of the experimental setup.
The measured $\langle \beta \rangle$ values are somewhat smaller for 500-pulse patterns with less attenuation 
or single pulse patterns with no attenuation, consistent with some disturbance of the sample structure by the incident beam.
For the single-pulse data, the energy density deposited is higher for Ni$_2$Pd$_2$P 
($\sim 75$ eV or $\sim 95$ eV per atom per average pulse)
than for Ga owing to its shorter absorption length, 
which agrees with the observed higher disturbance during single pulses \cite{Supplemental}.
Based on the maximum incident pulse energies in the attenuated 500-pulse sequences,
the threshold for longer time (inter-pulse) sample disturbance effects occurs at
about 1 eV per illuminated atom absorbed energy.
The dataset for B$_2$O$_3$ indicates that this threshold is also about 1 eV per atom in this ``light'' material,
corresponding to a higher incident fluence because of its longer absorption length.
Based on the observed melting threshold for Ni$_2$Pd$_2$P,
we estimate that the deposited pulse energy is thermalized in a volume 50 times larger than the
illuminated volume,
consistent with the expected spread of the electron/photon/phonon cascade
generated by x-ray absorption \cite{Supplemental},
resulting in a maximum temperature rise of $0.02 ~{\rm eV}/ 3 k_B \approx 80$~K.

These measurements allow an analysis of the feasibility of femtosecond XPCS experiments
using the split-pulse technique \cite{First_Experiments, GruebelNIM2007, GuttOE2009}.
The figure of merit for such measurements is the signal-to-noise ratio for $\beta$,
which for low $\bar{k}$ is determined by fluctuations in the small number of $k = 2$ events 
and can be  expressed as \cite{Supplemental}
$\beta / \sigma_\beta = \beta \bar{k} [n_{pix} n_{patt} / 2 (1+\beta)]^{1/2}$.
This expression agrees well with the observed accuracy of $\langle \beta \rangle$
given in Table I, indicating that other potential experimental contributions to the uncertainty
are negligible.
If we use a detector capable of recording $n_{pix} = 10^6$ pixels at the full LCLS repetition rate of 120 Hz,
to give $n_{patt} = 4 \times 10^4$ in six minutes per delay time,
and extrapolate the measured relationships between $\bar{k}$ and $I_0$ \cite{Supplemental},
the pulse energies needed to give a signal-to-noise ratio of 5
sufficient for XPCS 
are $I_0 = 6, 0.3,~{\rm or}~0.13 \times 10^8$ photons per pulse,
corresponding to sample temperature rises of 3 to 7~K,
for B$_2$O$_3$, Ga, or Ni$_2$Pd$_2$P,
respectively.
These temperature rises are significantly below the 80~K threshold
for structural disturbance by diffusive atomic rearrangement that we observed for the glass samples,
indicating that even more sensitive samples and processes can be studied.

The new analysis technique presented here allows the accurate
determination of speckle contrast 
using the weak patterns obtained under conditions
in which the x-ray pulses from a free electron laser
do not disturb the sample dynamics.
We observe high contrast factors in x-ray speckle patterns from liquids and glasses
obtained with ultrafast pulses,
demonstrating that XPCS can be used to observe their 
atomic-scale dynamics on times scales down to the femtosecond range.
We find that the atomic motions associated with x-ray damage
occur on time scales shorter than the x-ray pulse width used here
for deposited energy densities greater than about 50 eV per illuminated atom \cite{Supplemental}.
This provides guidance for design of studies in which single-pulse speckle patterns
are analyzed to obtain atomic-scale structural information.

Special thanks go 
to T. Hufnagel for the use of his metallic glass laboratory
and to A. Zholents for insightful comments.
X-ray experiments were carried out at LCLS at SLAC National Accelerator Laboratory, 
an Office of Science User Facility 
operated for the U.S. Department of Energy (DOE) Office of Science 
by Stanford University. 
Work at Argonne National Laboratory
supported by the DOE Office of Basic Energy Sciences 
under contract DE-AC02-06CH11357.  
Supporting electron microscopy was accomplished at 
the Electron Microscopy Center for Materials Research
at Argonne, 
a DOE Office of Science User Facility. 
The work of KFL on this project was supported by the U.S. DOE Office of Science, Office of Basic Energy Sciences under DE-FG02-03ER46037.

\end{document}